Technical note

# Daily milk yield correction factors: what are they?

Xiao-Lin Wu,[1,2] George Wiggans,[1] H. Duane Norman,[1] Asha M. Miles,[3] Curt Van Tassell,[3] Ransom L. Baldwin VI,[3] Javier Burchard,[1] and Joao Durr[1]

[1] Council on Dairy Cattle Breeding, Bowie, MD 20716, USA.
[2] Department of Animal and Dairy Sciences, University of Wisconsin, Madison, WI 53706, USA.
[3] USDA, Agricultural Research Service, Animal Genomics and Improvement Laboratory, Beltsville, MD 20705-2350, USA.

**Abstract:**

Cost-effective milking plans have been used to supplement the standard supervised four-weekly testing scheme since the 1960s. Cows are typically milked two or more times on a test day, but not all these milkings are sampled and weighed. Statistical methods have been proposed to estimate daily yields in dairy cows, centering on various yield correction factors in two broad categories, additive correction factors (ACF) and multiplicative correction factors (MCF). The initial approach estimated a test-day yield with doubled morning (AM) or evening (PM) yield in the AM-PM milking plans, assuming equal AM and PM milking intervals. However, AM and PM milking intervals can vary, and milk secretion rates may be different between day and night. ACF are evaluated by the average differences between AM and PM milk yield for various milking interval classes (MIC), coupled with other categorical variables. We show that an ACF model is equivalent to a regression model of daily yield on categorical regressor variables, and a continuous variable for AM or PM yield with a fixed regression coefficient of "2.0". Similarly, a linear regression model can be implemented as an ACF model with the regression coefficient for AM or PM yield estimated from the data. The linear regression models improved the accuracy of the estimates compared to the ACF models. MCF are ratio of daily yield to yield from single milkings, but their statistical interpretations vary. Overall, MCF were more accurate for estimating daily milk yield than ACF. MCF have biological and statistical challenges. Systematic biases occurred when ACF or MCF were computed on discretized MIC, leading to a loss of accuracy. An exponential regression model was proposed as an alternative model for estimating daily milk yield, which improved the accuracy in the present study. Characterization of ACF and MCF showed how ACF and MCF improved the accuracy compared to doubling AM or PM yield as the daily milk yield. The latter was approximately taken with equal AM and PM milkings only. The methods were explicitly described to estimate daily milk yield in AM and PM milking plans. Still, the principles are generally applicable to cows milked more than two times a day, and they apply similarly to the estimation of daily fat and protein yields with some necessary modifications.

**Keywords:** dairy cattle, daily milk yield curve, exponential function, milking interval, statistical models, yield correction factors

**Correspondence:** nick.wu@uscdcb.com

Accurate milking data are essential for herd management and genetic improvement in dairy cattle. Cows are typically milked two or more times on a test day, but not all these milkings are sampled and weighed. This practice started to supplement the standard supervised four-weekly testing scheme in the 1960s, motivated by reducing the visits by a National Dairy Herd Information Association (**DHIA**) supervisor and lowering the costs to the dairyman (Putnam and Gilmore, 1968). The initial AM-PM milking plan alternately sampled the morning (**AM**) or evening (**PM**) milking on test day throughout the lactation. Daily yield (milk, fat, and protein) was estimated by two times the yield from single milkings on each test day, assuming equal AM and PM milking intervals (Porzio, 1953). Let $x_{ij}$ be a known AM or PM yield on a test date, where $j = 1$ for AM milking or $j = 2$ for PM milking. Then, the total test-day yield is estimated by:

$$\hat{y}_{ij} = 2x_{ij} \qquad (1)$$

Mathematically, doubling AM or PM yield is equivalent to assuming a fixed multiplicative factor (or regression coefficient) of "2" for all cows, regardless of the actual milking intervals and milk secretion rates.

Various methods have been proposed to estimate daily yields, mainly to deal with varied milking intervals (See the graphical abstract). The landmark developments date to the 1980s and 1990s, centering on yield correction factors in two broad categories, additive correction factors (**ACF**) and multiplicative correction factors (**MCF**). In AM-PM plans, ACF provide additive adjustments to two times AM or PM milk yield (denoted by $x_{ijk}$) as the estimated daily yield, computed specifically for each milking interval class (MIC), say $k$. That is,

$$\hat{y}_{ijk} = \Delta_{jk} + 2x_{ijk} \qquad (2)$$

Here, $\Delta_{jk}$ represents an ACF for milking interval $k$ of milking $j$, and $\hat{y}_{ijk}$ is the estimated daily yield for cow $i$.

By noting that a daily milk yield equals the sum of the AM and PM milk yield, the following two equations can be derived for known AM and PM milk yield, respectively:

$$\Delta_{2k} = x_{i1k} - x_{i2k} \qquad (3)$$

$$\Delta_{1k} = x_{i2k} - x_{i1k} \qquad (4)$$

Thus, ACF are evaluated by the population averages of the differences between the AM and PM milk yield, coupled with other categorical variables whenever applicable (Everett and Wadell, 1970a). Everett and Wadell (1970a) have shown that the difference between AM and PM yields is a function of milking interval and days in milk (**DIM**). Significant variables affecting such differences vary with cattle breeds, including months of lactation, herd production level, age classes, MIC, and their interactions (Everett and Wadell, 1970b). A typical ACF model is a factorial (or analysis of variance) model with the difference between AM and PM yield as the response variable. The factors include, for example, MIC and lactation months.

Adding equation (3) to equation (4), we see that the sum of AM and PM ACF specific to each MIC equals is zero:

$$\Delta_{1k} + \Delta_{2k} = 0 \qquad (5)$$

A general form of an equivalent ACF model with daily yield as the response variable is the following:

$$y_{ijk} = f(\boldsymbol{\theta}) + bx_{ijk} + \epsilon_{ijk} \tag{6}$$

where $f(\boldsymbol{\theta})$ is a function with discrete variables, $b \equiv 2$ is the regression coefficient for AM or PM yield, and $\epsilon_{ijk}$ is a residual. Then, ACF are evaluated locally as the expected (**E**) values of $f(\boldsymbol{\theta})$ for each MIC:

$$\Delta_{jk} = E(f(\boldsymbol{\theta}; j, k)) \tag{7}$$

Similarly, a linear regression (**LR**) model can be implemented as an ACF model in which $f(\boldsymbol{\theta})$ is a linear function involving, say milking interval and DIM, and the regression coefficient for AM or PM yield is estimated from the data. Let $y_{ij}$ be a daily yield for cow $i$ on milking $j$, $t_{ij}$ be a milking interval time, $x_{ij}$ be a yield from a single milking of this animal for milking $j$, and $d_{ij}$ is the DIM for the test date when $x_{ij}$ is sampled and weighed. Then, the LR model involving the above variables alone is the following:

$$y_{ij} = \alpha_j + \beta t_{ij} + \gamma(d_{ij} - d_0) + bx_{ij} + \epsilon_{ij} \tag{8}$$

Here, $f(\boldsymbol{\theta}) = \alpha_j + \beta t_{ij} + \gamma(d_{ij} - d_0)$, $\alpha_j$ is an overall mean specific to milking $j$, and $\beta$, $\gamma$, and $b$ are common regression coefficients for milking interval, DIM, and single milking (AM or PM) yield, respectively. Note that LR models can be defined with varying complexity by adding or dropping regression variables as appropriate (Liu et al., 2000).

As a typical linear regression approach, daily yield is estimated directly given the estimated model parameters in (8):

$$\hat{y}_{ij} = \hat{\alpha}_j + \hat{\beta} t_{ij} + \hat{\gamma}(d_{ij} - d_0) + \hat{b}x_{ij} \tag{9}$$

Alternatively, ACF can be obtained by evaluating the expected values of $f(\boldsymbol{\theta})$ for discretized MIC locally. Assume that $E(d_{ij} - d_0) = 0$. Let there be $k = 1, \ldots, K$ classes for AM (or PM) milking. Then, ACF are computed for $2 \times K$ classes.

$$\Delta_j^{(k)} = E\left(\alpha_j + \beta t_{ij}^{(k)}\right) = \alpha_j + \beta E\left(\bar{t}_j^{(k)} + \left(t_{ij}^{(k)} - \bar{t}_j^{(k)}\right)\right) = \alpha_j + \beta \bar{t}_j^{(k)} + \beta E\left(t_{ij}^{(k)} - \bar{t}_j^{(k)}\right)$$

$$= \alpha_j + \beta \bar{t}_j^{(k)} \tag{10}$$

The above holds assuming $E\left(t_{ij}^{(k)} - \bar{t}_j^{(k)}\right) = E\left(t_{ij}^{(k)}\right) - \bar{t}_j^{(k)} = 0$, where $E(\bar{t}_{jk}) = \bar{t}_j^{(k)}$ is a midpoint of each MIC. Throughout this paper, we use a superscript "(k)" to indicate a discretized MIC because it is not a variable index in the data model. In contrast, a subscript index "k" is reserved for an index for a categorical variable in the data model. Then, daily yield is computed to be the sum of $\hat{b}x_{ij}$ and the computed ACF, $\Delta_j^{(k)}$, specific to a MIC. That is,

$$\hat{y}_{ij} = \hat{\Delta}_j^{(k)} + \hat{b}x_{ij} \tag{11}$$

Within each MIC, we show that the following holds in a joint analysis of AM and PM milking records, assuming a common regression coefficient for AM or PM yield.

$$\Delta_1^{(k)} + \Delta_2^{(k)} = (2 - b)\bar{y}^{(k)} \tag{12}$$

where $\bar{y}^{(k)}$ is the average daily milking yield for all the cows in MIC $k$. As a special case, if we force $b = 0$, the above relationship (12) is reduced to (5). Then, the LR model coincided with the an ACF model yet with continues variables for milking interval and DIM.

Multiplicative correction factors (**MCF**), also referred to as ratio factors, are ratios of daily yield to yield from single milkings, computed for various MIC (e.g., Shook et al., 1980; DeLorenzo and Wiggans, 1986; Wiggans, 1986). MCF replace the fixed multiplicative factor in (1) with unknown ratios for various MIC and estimated them from the data. Denote AMP and PMP for bulk AM and PM yield, respectively, and AMP + PMP gives the test-day yield. Then, the AM and PM multiplicative correction factors, denoted by $F_1$ and $F_2$, respectively, are defined as follows (Shook et al., 1986):

$$F_1 = \frac{AMP+PMP}{AMP} \tag{13}$$

$$F_2 = \frac{AMP+PMP}{PMP} \tag{14}$$

Confined to MIC $k$, we show the following relationship holds based on equations (13) and (14):

$$F_{1k}^{-1} + F_{2k}^{-1} = 1 \tag{15}$$

The above brings convenience to computing. For example, given the computed PM MCF ($F_{2k}$), AM MCF can be obtained indirectly as follows:

$$F_{1k} = (1 - F_{2k}^{-1})^{-1} = \frac{F_{2k}}{F_{2k}-1} \tag{16}$$

In practice, Shook et al. (1986) employed a quadratic regression of PM portion of daily yield on milking interval time to obtain smoothed estimates of MCF and dealt with MIC having no or insufficient milking records.

DeLorenzo and Wiggans (1986) proposed an LR model without intercept to derive MCF for cows milked twice a day. They assumed heterogeneous means and variances and fitted separate LR models for different MIC. Let $x_{ijk}$ be a yield from single milking $j$ of cow $i$ sampled on milking interval $k$, and $y_{ijk}$ be the corresponding daily yield. Then, an LR model defined for each MIC is the following:

$$y_{ijk} = F_{jk} x_{ijk} + \varepsilon_{ijk} \tag{17}$$

In the above, the regression coefficient $F_{jk}$ coincide by definition with the MCF specific to each MIC. Assuming $E(\varepsilon_{ijk}) = 0$, MCF for MIC k is also computed by:

$$F_{jk} = \frac{E(y_{ijk})}{E(x_{ijk})} == \frac{\frac{1}{n}\sum_{i=1}^{n} y_{ijk}}{\frac{1}{n}\sum_{i=1}^{n} x_{ijk}} = \frac{\sum_{i=1}^{n} y_{ijk}}{\sum_{i=1}^{n} x_{ijk}} \tag{18}$$

The above agrees with the MCF described by Shook et al. (1980) because $\sum_{i=1}^{n} x_{ijk}$ corresponds to $AMP_k$ or $PMP_k$ and $\sum_{i=1}^{n} y_{ijk}$ corresponds to $AMP_k + PMP_k$. Similar to Shook et al. (1980), yet somewhat differently, DeLorenzo and Wiggans (1986) proposed a linear smoothing by regressing the reciprocals of computed AM or PM factors on milking interval time to obtain smoothed MCF.

A covariate for DIM can also be included to account for the variation of the lactation curve:

$$y_{ijk} = F_{jk} x_{ijk} + \gamma_{jk}(d_{ijk} - d_0) + \varepsilon_{ijk} \tag{19}$$

where $\gamma_{jk}$ is the regression coefficient of DIM, and $d_0$ is a constant value.

Let $d_0 = E(d_{ijk})$ such that $E(d_{ijk} - d_0) = 0$, and let $E(\varepsilon_{ijk}) = 0$. Then, the computed yield factors based on the model (18) remain the same as (18). Here, $d_0$ is taken to be different for each MIC. Otherwise, we have $E(d_{ijk} - d_0) \neq 0$ and MCF are obtained as follows:

$$F_{jk} = \frac{E(y_{ijk} - \gamma_{jk}(d_{ijk} - d_0))}{E(x_{ijk})} = \frac{E(y_{ijk}) - \gamma_{jk} E(d_{ijk} - d_0)}{E(x_{ijk})}$$

$$= \frac{E(y_{ijk})}{E(x_{ijk})} - \gamma_{jk} \frac{E(d_{ijk} - d_0)}{E(x_{ijk})} \quad (20)$$

In the above, the second term in the right-hand side, $\gamma_j \frac{E(d_{ijk} - d)}{E(x_{ijk})}$, represents a bias yet to be adjusted. Alternatively, daily yield is adjusted by:

$$\hat{y}_{ijk} = F_{jk} x_{ijk} + \hat{\gamma}_{jk}(\bar{d}_{jk} - d_0) \frac{x_{ijk}}{\bar{x}_{jk}} \quad (21)$$

where $\bar{d}_{jk}$ and $\bar{x}_{jk}$ are the means of DIM and the yield from single milking (AM or PM), respectively, pertaining to MIC $k$ of milking $j$. Note that the adjustment in (21) is different from that in DeLorenzo and Wiggans (1986). We hold that the adjustment (21) is analytically precise.

Wiggans (1986) proposed deriving yield factors for cows milked three times a day through regression AM or PM proportion of daily yield ($\frac{x_{ij}}{y_{ij}}$) on milk interval ($t_{ij}$), as follows:

$$\frac{x_{ij}}{y_{ij}} = \alpha_j + \beta t_{ij} + \epsilon_{ij} \quad (22)$$

The same modeling strategies were used by Shook et al. (1980) and DeLorenzo and Wiggans (1986) because they had the same or similar ratio response variable in the linear or quadratic smooth functions. The difference is that the Wiggans (1986) model is a one-step LR fitted on individual milking data, but the methods by Shook et al. (1980) and DeLorenzo and Wiggans (1986) represented two-stage modeling. Another difference was that Wiggans (1986) assumed heterogenous intercepts and a common regression coefficient for AM and PM milkings. In contrast, the intercepts and regression coefficients were different for AM and PM milkings in the models by Shook et al. (1980) and DeLorenzo and Wiggans (1986). The Wiggans (1986) model also applies to cows milked more than three times and milked twice a day. In the latter case, the model is subject to the violation of linearity with a longer milking interval (Schmidt, 1960).

By taking the expected value on both sides of equation (22), and letting $E(\epsilon_{ij}) = 0$, we have:

$$E\left(\frac{x_{ij}}{y_{ij}}\right) = \alpha_j + \beta E(t_{ij}) \quad (23)$$

Here, $E(t_{ij}) = \bar{t}_j^{(k)}$ is evaluated locally as the midpoint of each MIC for milking $j$. Then, MCF are obtained as:

$$F_j^{(k)} = E\left(\frac{y_{ij}^{(k)}}{x_{ij}^{(k)}}\right) = \left(E\left(\frac{x_{ij}^{(k)}}{y_{ij}^{(k)}}\right)\right)^{-1}$$

$$= \frac{1}{\hat{\alpha}_j + \hat{\beta} \bar{t}_j^{(k)}} \quad (24)$$

Extending to having weighted $l'$ out of $L$ milkings, MCF are computed as follows:

$$F_{j \in l'}^{(k)} = E\left(\frac{y_{ij}^{(k)}}{\sum_{j \in l'} x_{ij}^{(k)}}\right) = \left(E\left(\frac{\sum_{j \in l'} x_{ij}^{(k)}}{y_{ij}^{(k)}}\right)\right)^{-1} = \frac{1}{\sum_{j \in l'} \hat{\alpha}_j + \hat{\beta} \sum_{j \in l'} \bar{t}_j^{(k)}} \quad (25)$$

When $E\left(d_{ij}^{(k)} - d_0\right) = 0$, MCF take the same forms as in (24) and (25). Otherwise, they need to be adjusted. For example, the formula (23) accounting for DIM then becomes:

$$F_j^{(k)} = \frac{1}{\hat{\alpha}_j + \hat{\beta}\bar{t}_j^{(k)} + \hat{\gamma}\left(\bar{d}_j^{(k)} - d_0\right)} \tag{26}$$

where $\bar{d}_j^{(k)}$ is the average DIM for records pertaining to MIC $k$ of milking $j$. Note that the adjustment needs to be made directly on the computed MCF because there is no simple adjustment term that is independent of the computed MCF.

The statistical interpretations of MCF vary slightly. Firstly, according to DeLorenzo and Wiggans (1986), an MCF is a regression coefficient specific to each MIC, defined in (17). Secondly, an MCF is a ratio of the expected value of daily yield to the expected value of yield from single milkings, computed for each MIC (Shook et al., 1980; DeLorenzo and Wiggans, 1986), defined in (18). Thirdly, an MCF is the expected value of the ratio of daily yield to a single milking yield according to the Wiggans (1986) model, defined in (24). Note that the third interpretation can also be derived from the regression smoothing models by Shook et al. (1980) and DeLorenzo and Wiggans (1986), because they fitted the same or similar ratio variable in the linear or quadratic smoothing models.

The three forms of MCF represent different strategies or formulations for estimating ratios of daily yield to yield from single milkings. Hence, they correspond to each other approximately. For example, the form in (24) agrees with (18) approximately if we apply the first-order Taylor approximation to (24). That is,

$$E\left(\frac{y_{ij}}{x_{ij}}\right) \approx \frac{E(y_{ij})}{E(x_{ij})} \tag{27}$$

Multiplicative correction factor models are statistically challenged by the well-known "the ratio problem" because they have a ratio variable (e.g., AM or PM proportion of daily yield) as the dependent variable in the data density (Wiggans, 1986) or the smoothing functions (Shook et al., 1980; DeLorenzo and Wiggans, 1986). The consequences included possible biases in two aspects: omitted variable bias and measurement error bias (Lien et al., 2017). The former happens because main model effects are missing, which can be seen if the model is re-arranged by multiplying the denominator variable to both sides of the equation. The latter occurs when there are measurement errors with the denominator variable of the response. Besides that, the MCF models postulated a rational function between daily milky yield and milking, in which the numerator is one, and the denominator is a linear function (DeLorenzo and Wiggans, 1986; Wiggans, 1986) or a quadratic function (Shook et al., 1980) of milking interval. Yet, no biological evidence has been available to support a rational function as a daily milk curve.

Here, we propose an alternative model by taking the logarithm of daily to single milking (say AM or PM) yield ratio as a response variable. That is,

$$\log\left(\frac{y_{ij}}{x_{ij}}\right) = \alpha_j + \beta t_{ij} + \epsilon_{ij} \tag{28}$$

where $\frac{y_{ij}}{x_{ij}}$ is a ratio of daily yield to single milking (say AM or PM) yield. With some re-arrangements, equation (30) becomes:

$$\log(y_{ij}) = \alpha_j + \beta t_{ij} + b\log(x_{ij}) + \epsilon_{ij} \tag{29}$$

Here, $\log(y_{ij})$ is the response variable, $\log(x_{ij})$ and $t_{ij}$ are the dependent variables (i.e., main effects), and $b = 1$ is a constant regression coefficient for $\log(x_{ij})$. In the present study, however, we relax the restriction for $b = 1$ in (31) and estimate it from the data. Then, taking the exponential on both sides of equation (29) gives:

$$y_{ij} = x_{ij}^b \, e^{(\alpha_j + \beta t_{ij} + \epsilon_{ij})} \qquad (30)$$

The above is recognized as an exponential regression model. The model parameters can be conveniently estimated with data fitted on the linear logarithm model (29). Daily yield is calculated given the model parameter estimates ($\hat{b}, \hat{\alpha}_j, \hat{\beta}$) and observed partial (AM or PM) yield and milking interval time.

To derive MCF, we first take expected values on both sides of equation (30). Then, we applied the second Taylor approximation by noting that $E(log(z)) \approx log(E(z)) - \frac{V(z)}{2E(z)^2}$, where z is a random variable. Hence, we have:

$$log(E(y_i)) = \alpha_j + \beta \left(E(t_{ij})\right) + b\,log\left(E(x_{ij})\right) + \left(\frac{V(y_{ij})}{2E(y_{ij})^2} - b\frac{V(x_{ij})}{2E(x_{ij})^2}\right) \qquad (31)$$

Next, taking the exponential on both sides of equation (31), with some re-arrangements, gives:

$$E(y_{ij}) = \rho E(x_{ij})^b e^{\{\alpha_j + \beta E(t_{ij})\}} \qquad (32)$$

where $\rho = e^{\frac{1}{2}\left(V(y_{ij})E(y_{ij})^{-2} - bV(x_{ij})E(x_{ij})^{-2}\right)}$. MCF are derived by taking the expected values of (32) locally for each MIC, say $k$, and dividing both sides of the equation by $E\left(x_{ij}^{(k)}\right)$. That is,

$$F_j^{(k)} = \frac{E\left(y_{ij}^{(k)}\right)}{E\left(x_{ij}^{(k)}\right)} = \rho_j^{(k)} E\left(x_{ij}^{(k)}\right)^{b-1} e^{\alpha_j + \beta \bar{t}_j^{(k)}} \qquad (33)$$

where $\rho_j^{(k)} = e^{\frac{1}{2}\left(V(y_{ij}^{(k)})E(y_{ij}^{(k)})^{-2} - bV(x_{ij}^{(k)})E(x_{ij}^{(k)})^{-2}\right)}$, and $E\left(y_{ij}^{(k)}\right) = \bar{y}_j^{(k)}$ and $E\left(x_{ij}^{(k)}\right) = \bar{x}_j^{(k)}$ are the corresponding means for daily yield and AM (or PM) yield, respectively.

In the present study, various models were evaluated and compared using simulation datasets. Daily milk yields were simulated based on a modified Michael-Menten function (Klopcic et al., 2012). The Michaelis–Menten function is a well-known model for enzyme kinetics (Srinivasan, 2021). Klopcic et al. (2012) showed that a modified Michael-Menten function fitted daily milk yield curves very well. Daily milk yield curves were simulated for 3,000 cows (See graphical abstract), where the values for $y_{720}$ and $k$ were simulated from truncated normal (TN) distributions: $y_{720} \sim TN(12,2)$ and $k \sim TN(0.8,0.1)$. AM milking intervals were simulated following a truncated normal distribution with a mean equaling 12 hours and a standard deviation of 1.12 hours. PM milking intervals for the same cows were 24 hr. minus the AM milking intervals. Approximately 98.6% of the cows had AM (PM) milking intervals between 9 hr. and 15 hr. (See graphical abstract). Deviations due to days in milk and other system variables were ignored to simplify the discussion.

The performance of each model was evaluated based on mean squared errors (MSE) and accuracies of estimated daily milk yields obtained from ten-fold cross-validations, each replicated $M = 30$ times. In each replicate, 2,000 randomly selected cows were used for training and the remaining 1,000 animals were used for testing. The $R^2$ accuracy (Liu et al., 2000) was computed as follows, and averaged across 30 replicates for each method:

$$R^2 = \frac{\sigma^2}{\sigma^2 + MSE} \qquad (34)$$

Here, $\sigma^2$ was the phenotypic variance of true (simulated) test-day milk yield. To further infer the origin of errors, MSE was decomposed into variance ($Var(\hat{y}_i)$) and squared bias ($Bias^2(\hat{y}_i)$), where $\hat{y}_i$ was an estimate of daily milking yield for individual *i*. Note that biases are derived relative to actual test-day milk yields. But keep in mind that milking records are subject to further adjustment or standardization accounting for, e.g., number of milkings, length of lactation, age, and month at calving before genetic evaluation (McDaniel, 1973; Schutz and Norman, 1994; Norman et al., 1995). The latter is not covered in the present study.

The variances of estimated daily milk yields were all close to zero for these methods (Table 1), suggesting that they all had high precision for the estimates. Overall, the MCF models had slightly smaller squared biases (and MSE) and better accuracies than the ACF models (Table 1). The two ACF models, M1 and M2B, had the largest MSE and the lowest $R^2$ accuracy. For the MCF models, M6A and M6B performed (Wiggans, 1986) slightly better than M4 (Shook et al., 1980) and M5 (DeLorenzo and Wiggans, 1986). The latter two models, M4 and M5, performed similarly. Model M6B estimated daily milk yield based on MCF (Wiggans, 1986). The two exponential function models, M7A and M7B, had the smallest squared bias (and MSE) and the largest accuracies in the methods.

(**Table 1**)

Estimated model parameters for four models were obtained in one cross-validation replicate and shown in Table 2. The results also included linear regression fits and correlation coefficients between actual and estimated daily milk yields. The correlations were high for all the models, showing very slight differences. Yet, the fitted linear regressions between actual and estimated daily milk yields varied considerably between these models. The ACF model (M1) had larger intercepts than the MCF model. The LR model with continuous milking interval (M2A) had a significantly smaller intercept. The exponential regression model, M7A, had the smallest intercept, and the regression coefficient was close to 1. Hence, the ACF model M1 had the largest biases and the worst accuracies. The exponential model M7A had the least biases and the greatest accuracies. These results also suggested that correlations are not an appropriate measure of accuracy for estimating daily milk yields because biases are not considered.

(**Table 2**)

The impact of discretizing milking intervals on estimating daily milk yields was evaluated through four pairs of models. Each pair had the same model settings, except that daily milk yields were estimated with different strategies. The models labeled "A" (M3A, M3A, M6A, and M7A) estimated daily milk yields directly based on estimated model parameters. Instead, the models labeled "B" (M2B, M3B, M6B, and M7B) first computed ACF or MCF for discretized MIC after data fitting. Then, daily milk yields were estimated based on the computed ACF or MCF per discretized MIC. The models in group A consistently had smaller MSE and better accuracies than their counterparts in group B. We thus concluded that discretizing milking interval time led to increased biases and, therefore, loss of accuracies in estimated daily milk yield. Systematic biases arose from discretizing milking interval, which is analytically shown below. Consider the LR model M3A. Given the model parameters, it estimated daily milk yield as follows:

$$\hat{y}_{ij} = \hat{\alpha}_j + \hat{\beta} t_{ij} + \hat{b} x_{ij}$$
$$= \left(\hat{\alpha}_j + \hat{\beta}\bar{t}_j^{(k)}\right) + \hat{\beta}\left(t_{ij} - \bar{t}_j^{(k)}\right) + \hat{b} x_{ij} \qquad (35)$$

The estimated daily milk yield consisted of three components in (35). In contrast, the model M3B computed ACF for discretized MIC, and then daily milk yield was computed by the first and the third components in (35):

$$\hat{y}_{ij} = \left(\hat{\alpha}_j + \hat{\beta}\bar{t}_j^{(k)}\right) + \hat{b}x_{ij} \qquad (36)$$

where $\hat{\alpha}_j + \hat{\beta}\bar{t}_j^{(k)}$ corresponds to an ACF. Hence, the second term on the right-hand side of (35), $\hat{\beta}\left(t_{ij} - \bar{t}_j^{(k)}\right)$, represented a systematic bias which the model M3B ignored. A similar situation happened with the models M2A and M2B. For the exponential model M7B, the bias due to discretizing interval milking was quantified by $e^{\hat{\beta}\left(t_{ij}-\bar{t}_j^{(k)}\right)}$. Biases due to discretizing milking interval also existed with the model M6B (Wiggans, 1986). Nevertheless, the biases from discretizing milking intervals were relatively small in the present study.

Additive and multiplicative correction factors were characterized and compared in Figure 1. ACF computed from the two ACF models, M1 and M2B, were comparable except that ACF from the model M2B were smoothed. But they did not agree with ACF computed from the LR model M3B. The average difference of ACF per MIC between M2B and M3B was $1.402/2 = 0.701$. Analytically, we confirmed the results by deriving a similar average difference of ACF between these two models:

$$\bar{\Delta}_{M3B} - \bar{\Delta}_{M2B} = \frac{1}{K}\sum_{k=1}^{K}\left(\Delta_{M3B}^{(k)} - \Delta_{M2B}^{(k)}\right) = (2-b)\left(\frac{1}{2K}\sum_{j=1}^{2}\sum_{k=1}^{K}\bar{x}_j^{(k)}\right) \approx 0.699$$

where $b = 1.942$, and $\frac{1}{2K}\sum_{j=1}^{2}\sum_{k=1}^{K}\bar{x}_j^{(k)} = 12.05$. The sums of AM and PM ACF within MIC were all close to zero for the ACF model. For the LR model, the sums of AM and PM ACF within MIC was approximately 1.402. Analytically, the sum was estimated to be: $(2-\hat{b})\left(\frac{1}{2K}\sum_{j=1}^{2}\sum_{k=1}^{K}\bar{y}_j^{(k)}\right) = 1.398$ according to (12), where $\frac{1}{2K}\sum_{j=1}^{2}\sum_{k=1}^{K}\bar{y}_j^{(k)} = 24.10$.

Multiplicative correction factors were computed and compared between four models: M4 (Shook et al., 1980), M5 (DeLorenzo and Wiggans, 1986), M6B (Wiggans, 1986), and M7B. Overall, MCF computed from different models are highly comparable for milking intervals between 9 hr. and 14 hr., but they showed relatively larger differences out of this range (Figure 1). All the computed MCF provided estimates of the ratios of daily yield to yields from single milkings, though their precise statistical interpretations varied.

(**Figure 1**)

Estimating test-day yield by doubled AM or PM milk yield was approximately taken with equal AM or PM milking interval. Still, it was subject to large errors when AM or PM milking interval deviated from 12 hours. This conclusion was made based on two observed facts. Firstly, ACF from the model M1 and M2B were zero when AM and PM milking intervals were both approximately 12 hours, meaning that close to zero adjustments were added to two times AM or PM milk yield as the estimated daily milk yields. Secondly, MCF were all close to 2.0 with approximately 12-12 hr. equal AM and PM milking intervals. MCF became smaller or larger than 2.0 as AM or PM milking interval departure from 12-12 hours.

By noting $e \approx 2.718$, we show that the exponential function is also analogous to an exponential growth function:

$$y = x^b(1 + 1.718)^{t^*} \tag{37}$$

where $y_0 = x^b$ is the initial value, $r = 1.718$ is the rate of change, tuned by a time function as a linear function of milking interval and days in milk. The newly proposed model (M7A and M7B) postulated an exponential growth dynamics between daily milk yield and milking interval time. In reality, daily milk yield (including fat and solids-not-fat) was not linear with intervals beyond 12 hr. (Ragsdale et al., 1924; Bailey et al., 1955; Elliott and Brumby, 1955; Schmidt,1960). Early studies also supported exponential daily milk curves. For example, Brody (1945) showed milk yields and fat percentages for milking intervals between 1 and 36 hr., which empirically resembled an exponential function. In relation to the interval between the current milking and the previous milking, milk and component productions showed an exponential increase at the beginning and later leveled off to an asymptote, possibly result from cell degradation and milk present in the udder (Neal and Thornley, 1983).

In conclusion, we reviewed and evaluated ACF and MCF models concerning their statistical interpretations, model assumptions, and challenges. The exponential regression model provided an alternative tool for estimating daily milk yield with improved the accuracy. In a continuing effort, large-scaled high-resolution milking data are being collected for follow-up studies, jointly supported by the US Council on Dairy Cattle Breeding, the USDA Agricultural Genomics and Improvement Laboratories, and the National Dairy Herd Information Association.

Table 1. **Variance, squared bias (bias²), and mean squared error (MSE) of estimated daily milk yields using different models and strategies**

| Models | Variance | Bias² | MSE | Accuracy |
|---|---|---|---|---|
| M1 | 4.90E-04 | 0.486 | 0.486 | 0.968 |
| M2A | 7.40E-05 | 0.448 | 0.448 | 0.971 |
| M2B | 7.50E-05 | 0.480 | 0.480 | 0.968 |
| M3A | 7.40E-05 | 0.435 | 0.435 | 0.972 |
| M3B | 7.40E-05 | 0.465 | 0.465 | 0.970 |
| M4 | 4.60E-04 | 0.422 | 0.422 | 0.972 |
| M5 | 4.30E-04 | 0.420 | 0.421 | 0.972 |
| M6A | 5.90E-05 | 0.386 | 0.386 | 0.975 |
| M6B | 5.90E-05 | 0.417 | 0.417 | 0.973 |
| M7A | 5.90E-05 | 0.376 | 0.376 | 0.976 |
| M7B | 7.60E-05 | 0.385 | 0.385 | 0.975 |

[1] Models

M1: $y_{ijk} = \mu_{jk} + 2x_{ijk} + \epsilon_{ijk}$; $\Delta_{jk} = \hat{\mu}_{jk}$.

M2A: $y_{ij} = \alpha_j + \beta t_{ij} + 2x_{ij} + \epsilon_{ij}$; M2B: $y_{ij} = \alpha_j + \beta t_{ij} + 2x_{ij} + \epsilon_{ij}$; $\Delta_j^{(k)} = \hat{\alpha}_j + \hat{\beta}\bar{t}_j^{(k)}$.

M3A: $y_{ij} = \alpha_j + \beta t_{ij} + b x_{ij} + \epsilon_{ij}$; M3B: $y_{ij} = \alpha_j + \beta t_{ij} + b x_{ij} + \epsilon_{ij}$; $\Delta_j^{(k)} = \hat{\alpha}_j + \hat{\beta}\bar{t}_j^{(k)}$.

M4: $\frac{\sum_i x_{ij}^{(k)}}{\sum_i y_{ij}^{(k)}} = \alpha_j + \beta_{j1}\bar{t}_j^{(k)} + \beta_{j2}\left(\bar{t}_j^{(k)}\right)^2 + \epsilon_j$; $F_j^{(k)} = \frac{1}{\hat{\alpha}_j + \hat{\beta}_{j1}\bar{t}_j^{(k)} + \hat{\beta}_{j2}\left(\bar{t}_j^{(k)}\right)^2}$.

M5: $y_{ijk} = b_{jk}x_{ijk} + \varepsilon_{ijk}$; $\left(\hat{b}_{jk}\right)^{-1} = \alpha_j + \beta_j\bar{t}_j^{(k)} + \epsilon_j$; $F_j^{(k)} = \frac{1}{\hat{\alpha} + \hat{\beta}\bar{t}_j^{(k)}}$.

M6A: $\frac{x_{ij}}{y_{ij}} = \alpha_j + \beta t_{ij} + \epsilon_{ij}$; M6B: $\frac{x_{ij}}{y_{ij}} = \alpha_j + \beta t_{ij} + \epsilon_{ij}$; $F_j^{(k)} = \frac{1}{\hat{\alpha} + \hat{\beta}\bar{t}_j^{(k)}}$

M7A: $y_{ij} = x_{ij}^b e^{(\alpha_j + \beta t_{ij} + \epsilon_{ij})}$; M7B: $y_{ij} = x_{ij}^b e^{(\alpha_j + \beta t_{ij} + \epsilon_{ij})}$; $F_j^{(k)} = \rho_j^{(k)}\left(\bar{x}_j^{(k)}\right)^{b-1} e^{\hat{\alpha}_j + \hat{\beta}\bar{t}_j^{(k)}}$.

[2] Variable notations

$y_{ij}$ ($y_{ijk}$) = daily milk yield for cow $i$; $x_{ij}$ ($x_{ijk}$) = milk yield of cow $i$ sampled from milking $j$ for $j = 1$ (AM milking) or 2 (PM milking); k is an index for milking interval classes when applicable; $\alpha_j$ = intercept defined for milking $j$; $\beta$ = homogenous regression coefficient of milking interval, $b$ = regression coefficient of AM (or PM) milk yield, and $\epsilon_{ij}$ = residual. $\Delta_j^{(k)}$ = additive correction factor for interval class $k$ of milking $j$; $F_j^{(k)}$ = multiplicative correction factor for milking interval $k$ of milking $j$; $\bar{t}_j^{(k)}$ = midpoint of milking interval $k$; $\rho_j^{(k)}$: see equation (38).

**Table 2.** Estimated model parameters, linear regression fits and corrections between actual ($y$) and estimated ($\hat{y}$) daily milk yields obtained from four statistical models. [1,2,3,4]

| Statistical models | Model parameters | | | | Linear regression fit | Correlation |
|---|---|---|---|---|---|---|
| | $\alpha_{AM}$ | $\alpha_{PM}$ | $\beta$ | $b$ | | |
| M2A | 14.17 | 14.19 | -1.182 | 2.0 | AM: $y = 0.822 + 0.966\,\hat{y}$ | 0.985 |
| | (0.095) | (0.095) | (0.008) | --- | PM: $y = 0.580 + 0.976\,\hat{y}$ | 0.985 |
| M3A | 14.46 | 14.48 | -1.147 | 1.942 | AM: $y = 0.123 + 0.995\,\hat{y}$ | 0.986 |
| | (0.096) | (0.096) | (0.008) | (0.004) | PM: $y = -0.126 + 1.005\,\hat{y}$ | 0.985 |
| M6A | 0.208 | 0.208 | 0.024 | --- | AM: $y = 0.684 + 0.972\,\hat{y}$ | 0.986 |
| | (0.002) | (0.002) | (<0.001) | --- | PM: $y = 0.577 + 0.976\,\hat{y}$ | 0.986 |
| M7A | 1.324 | 1.324 | -0.048 | 0.977 | AM: $y = 0.102 + 0.996\,\hat{y}$ | 0.988 |
| | (0.005) | (0.005) | (<0.001) | (0.002) | PM: $y = -0.009 + 1.001\,\hat{y}$ | 0.987 |

[1] M2A, M3A, M6A, M7A: See model specifications in Table 1.
[2] Each model assumed heterogeneous intercepts for AM and PM milking ($a_1$ and $\alpha_2$), respectively, and a common regression coefficient ($\beta$) for milking interval; $b$ = regression coefficient of single milking (AM or PM) yield.
[3] Numbers in paratheses are standard deviations of estimated model parameters.
[4] "---" = not applicable.

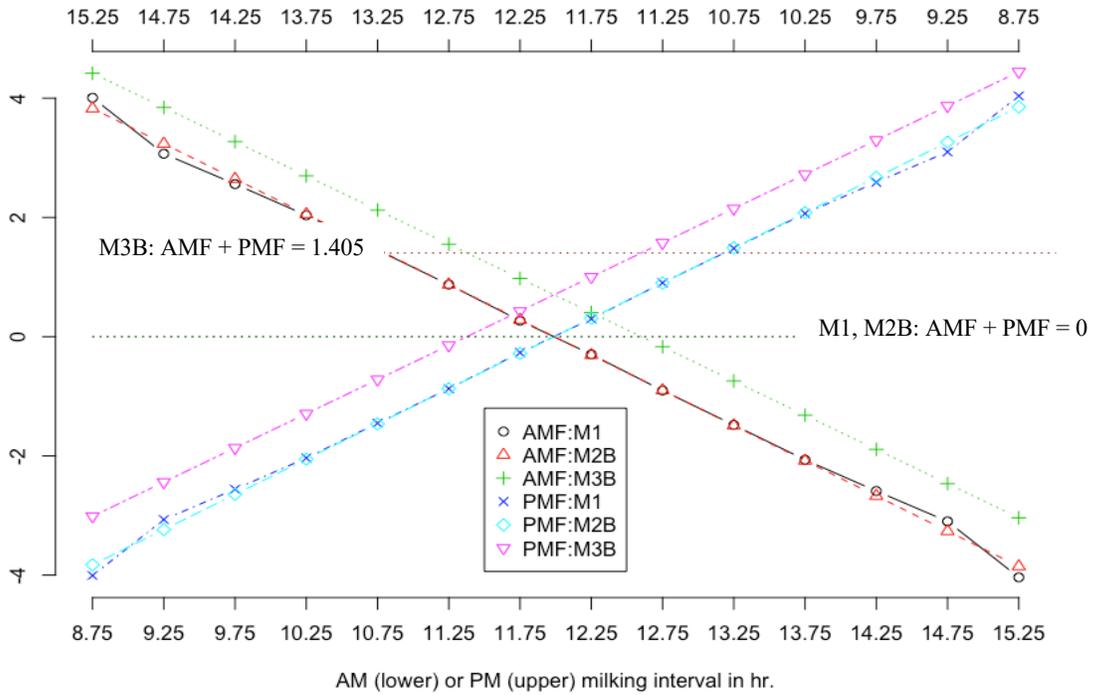

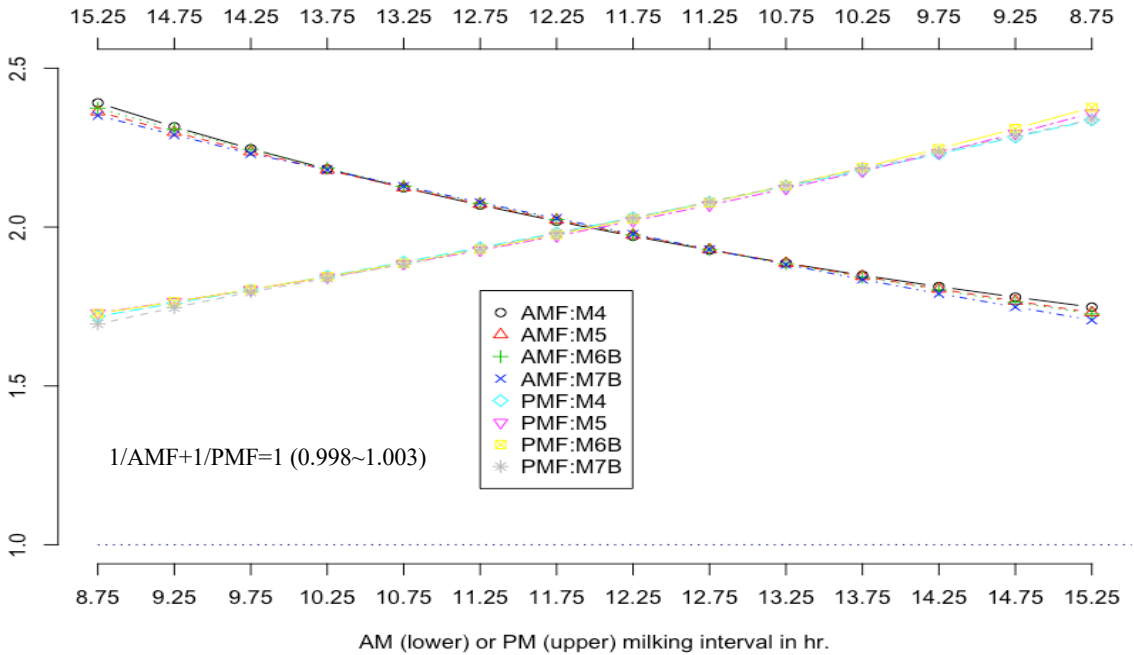

Figure 1. Comparing additive correction factors (upper) and multiplicative correction factors (lower) obtained using different methods.

M1, M2B, M3B, M4, M5, M6B, M7B, M8B: see model specifications in Table 1. AMF = morning correction factors; PMF = evening correction factors.